\documentclass[a4paper,11pt]{article}
\usepackage{pos}
\usepackage{float}
\usepackage{wrapfig}
\usepackage[capitalise]{cleveref}
\usepackage{subcaption}
\usepackage{enumitem}
\usepackage[left]{lineno}
\title{Ultra-High-Energy Transient Source Search using IceCube Neutrinos and Pierre Auger Photon Candidates}
 \ShortTitle{Ultra-High-Energy Transient Sources Search}

\author*[a]{Maxwell Nakos}

\affiliation[a]{University of Wisconsin-Madison}

\onbehalf{for the IceCube$^b$ and Pierre Auger Collaborations$^c$}
\affiliation[b]{The IceCube Collaboration (a complete list of authors can be found at the end of the proceedings)}
\affiliation[c]{Observatorio Pierre Auger, Av.\ San Mart{\'\i}n Norte 304, 5613 Malarg\"ue, Argentina\\
Full author list: {\rm\url{https://www.auger.org/archive/authors_icrc_2025.html}}}

\emailAdd{maxwell.nakos@icecube.wisc.edu}
\emailAdd{spokespersons@auger.org}

\abstract{The origins of ultra-high-energy particles remain one of the most profound mysteries in astrophysics. If nearby transient sources of ultra-high-energy particles exist, we might expect correlated emission of neutrinos and photons, arriving in close temporal and spatial coincidence. The IceCube Neutrino Observatory, located at the South Pole, is sensitive to neutrinos from TeV to EeV energies, while the Pierre Auger Observatory, in Argentina, detects cosmic rays and has the capability of observing ultra-high-energy photons using surface and fluorescence detectors. The ultra-high-energy photon candidates reported by the Auger Collaboration, though consistent with cosmic ray backgrounds, provide a compelling opportunity to search for correlated neutrino-photon events. In this contribution, we present the framework to search for ultra-high-energy transients by combining multi-flavour neutrino data from IceCube with photon candidates from the Auger Detector from 2011 to 2017. We will report the sensitivities that we expect to achieve from the search and the astrophysical implications of the possible outcomes.}

\ConferenceLogo{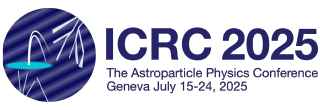}
\FullConference{39th International Cosmic Ray Conference (ICRC2025)\\
 15–24 July 2025\\
Geneva, Switzerland\\}

\begin{document}
\maketitle

\section{Introduction}
\begin{figure*}[h!]
    \centering
    \includegraphics[width=1\textwidth]{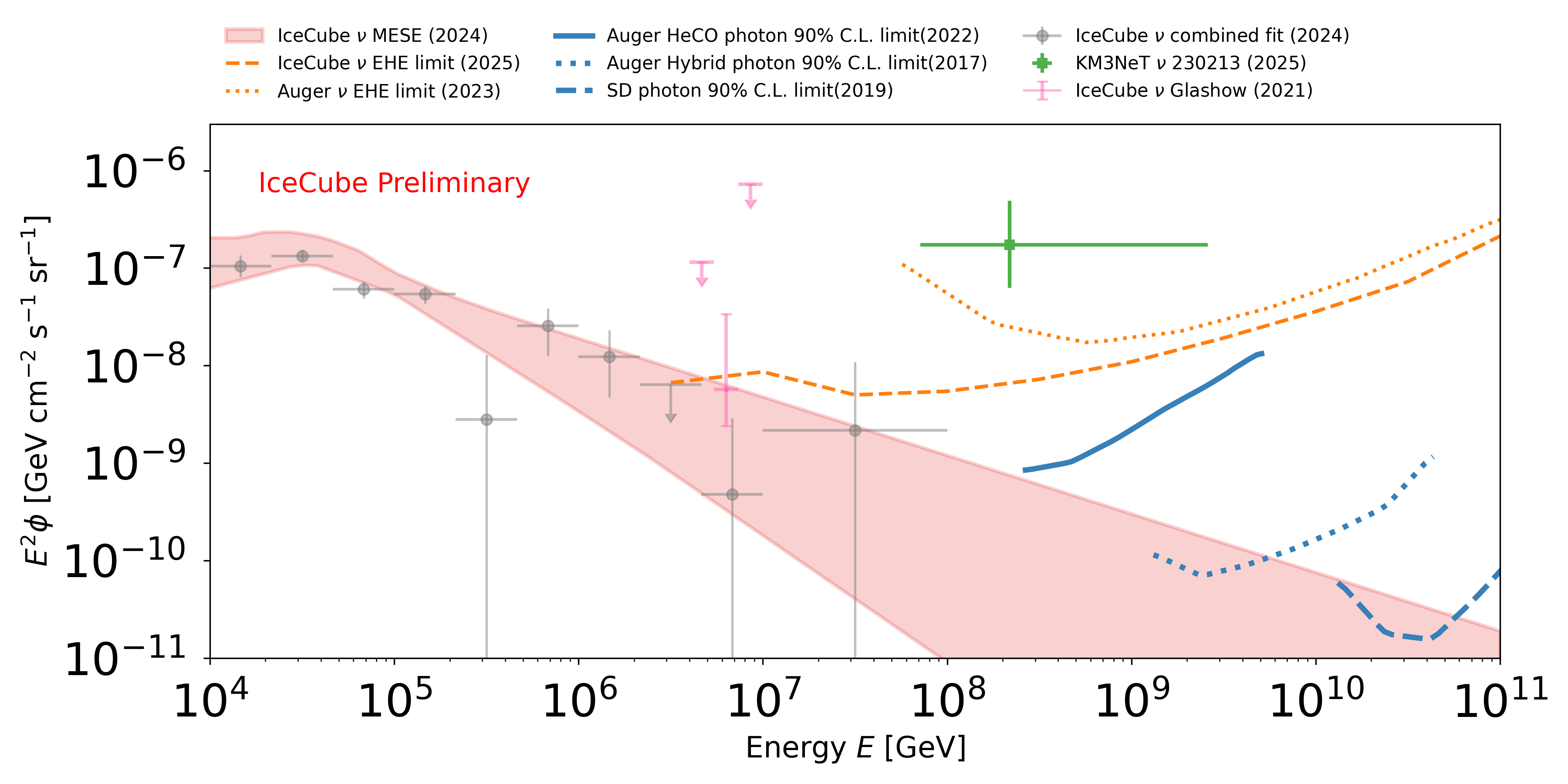}
    \caption{Multi-messenger diffuse flux measurements for neutrinos~\cite{combinedfit, glashow}. Upper limits for the ultra-high energy photons is included~\cite{photonlow, photon_mid, photon_high}. Upper limits are compared to the projection of the diffuse neutrino flux using measurements from the Medium Energy Starting Events (MESE) selection~\cite{mese}. Differential limits are shown for the high-energy neutrino flux set by IceCube and Auger Detector~\cite{EHE2025, nuupperlimit}. The measurement from the KM3-230213A neutrino is also shown~\cite{km3event}.}
    \label{fig:diffuse}
\end{figure*}

An enduring question in astrophysics is the origin of ultra-high-energy particles. One method for finding these sources is through multi-messenger astronomy. Proton-proton (pp) or proton-gamma (p$\gamma$) interactions in a nearby source could yield photons and neutrinos with energies greater than one EeV. We will search for ultra-high-energy sources using spatial and timing coincidences between two messengers. For a similarly sized detector to the Pierre Auger Observatory, accounting for photon interactions while propagating in the cosmic microwave background (CMB), we would expect to detect photons with energies greater than 10 EeV from a source similar to a low-luminosity gamma-ray burst (LLGRB) within 10 Mpc, after which the detection probability decreases~\cite{distance}. Coincident photons and neutrinos from nearby astrophysical sources should be detectable. \Cref{fig:diffuse} illustrates the diffuse landscape at energies above 10 TeV for neutrinos, UHE photons, and cosmic rays. Assuming neutrinos and photons are produced by the same hadronic interactions at the source, we would expect that the neutrino flux and UHE photon flux to be comparable. This is illustrated by the comparison of the upper limits on the diffuse UHE photon flux set by the Auger Detector~\cite{photonlow, photon_mid, photon_high} to the extrapolated diffuse neutrino flux measured by the Medium Energy Starting Events (MESE) selection in IceCube~\cite{mese}. In this analysis, we will be looking for photons and neutrinos produced by a transient flare from the same astrophysical source.

\subsection{IceCube Neutrino Observatory}
IceCube is a cubic kilometer neutrino observatory located at the geographic South Pole. It provides full-sky coverage with a greater than 99\% uptime \cite{instrumentation}. IceCube has performed many prior searches for transient sources. These searches have resulted in upper limits on the neutrino emission from high luminosity Gamma Ray Bursts (GRBs), such as GRB 221009A~\cite{grb}. Searches for coincident neutrinos from gravitational wave (GW) emission, such as during the O3 run, have not found significant emission~\cite{gw_search}. IceCube has found evidence for neutrino emission from the flaring blazar TXS 0506+056, which has coincident gamma-ray emission~\cite{txs}. 
\subsection{Pierre Auger Observatory}
The Pierre Auger Observatory is a cosmic ray detector composed of a surface detector (SD) and florescent detector (FD). The SD comprises of 1660 water-Cherenkov detectors distributed over an area of 3000 km$^2$. The FD is composed of 27 telescopes operated at 4 different sites to measure the longitudinal development of the air shower~\cite{pierre_bg}. In addition to detecting cosmic rays, the Pierre Auger Observatory is also capable of observing EeV photon showers. Using events detected by the FD in combination with the SD, photon candidates are selected using differences in muon fraction and the depth of the shower maximum compared to cosmic ray showers~\cite{photon}. 
\section{Analysis Method}
In this proposed analysis, we are searching for neutrinos from the direction of photon candidates. These candidates will have a similar treatment to previous IceCube searches of neutrinos from the same direction and time as gravitational waves~\cite{grav}. To accomplish this, we seek a significant neutrino detection in spatial and timing coincidence with the photon candidate while accounting for the uncertainty in photon reconstruction. This section details how photon candidates and IceCube neutrino event selections are incorporated into the analysis. Additionally, we outline the unbinned maximum likelihood method that will be employed in the analysis.
\subsection{Photon Candidates}
To perform this analysis, we utilize UHE photon candidates with energies greater than one EeV from the Auger Observatory as a prior for a source location, where the uncertainty in the directional reconstruction map is accounted for by a spatial probability density function (PDF). For hybrid events, the uncertainty in reconstructed direction is approximately one degree. While we sample these photons from the hybrid photon analysis~\cite{photon}, the Auger Collaboration has performed additional searches for UHE photons~\cite{sdonly, photonlow, photon_mid, photon_high}. \Cref{fig:exposure} shows the exposure for the hybrid photon selection, i.e. events detected by the FD in combination with the SD. 
\subsection{Neutrino Selections}
\begin{figure*}[h!]
    \centering
    \begin{subfigure}[t]{0.43\textwidth}
        \centering
        \includegraphics[width=1\textwidth]{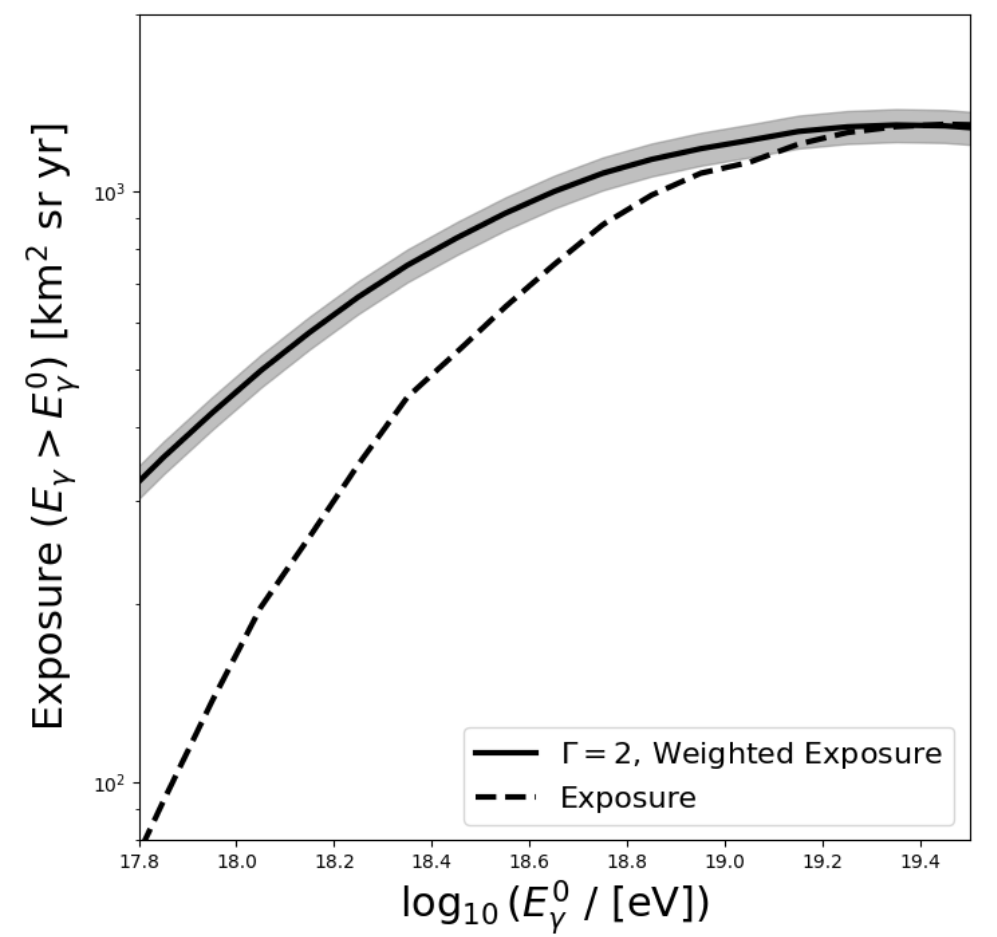}
        \caption{}
        \label{fig:exposure}
    \end{subfigure}
    ~ 
    \begin{subfigure}[t]{0.55\textwidth}
        \centering
        \includegraphics[width=1\textwidth]{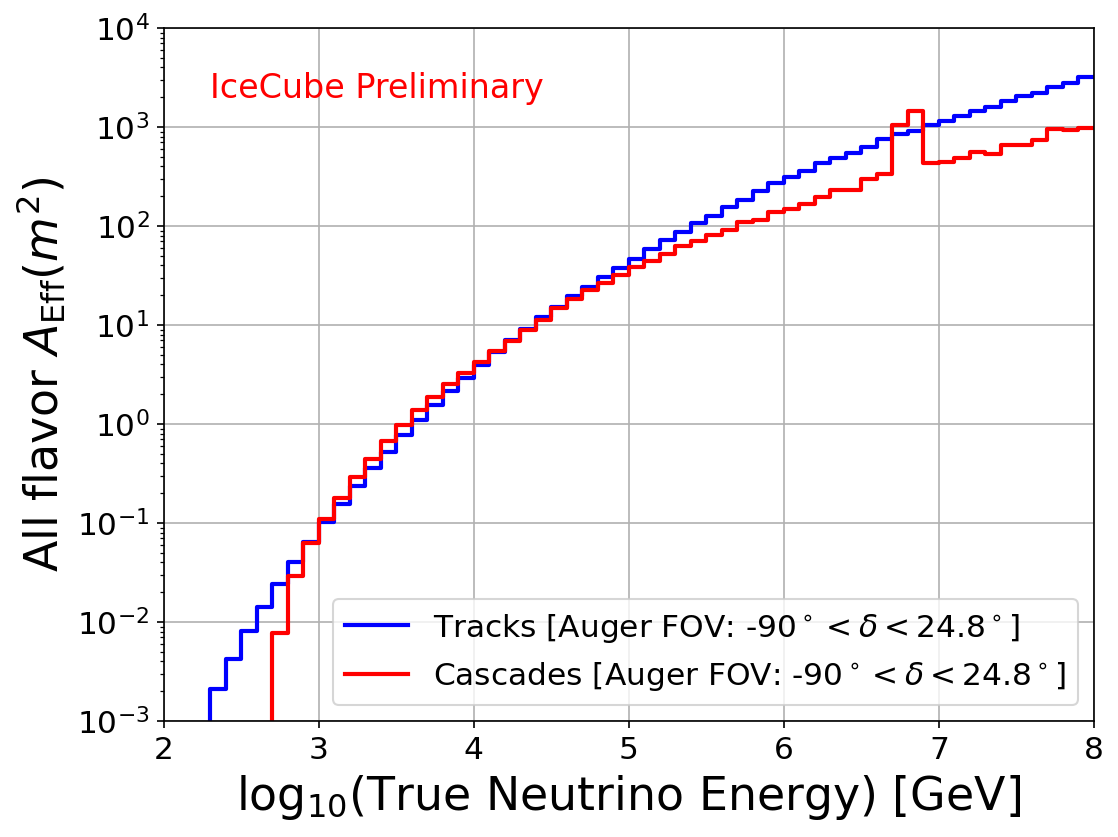}
        \caption{}
        \label{fig:effective area}
    \end{subfigure}%
    \caption{\textbf{(a)} The exposure for the photons analysis from \cite{photon}. The weighted exposure (solid line) assumes a power-law spectrum of $\gamma = 2$. The raw exposure (dashed line) is also shown for comparison.  \textbf{(b)}  The IceCube tracks~\cite{pstracks} and cascades~\cite{dnn} selection effective areas in the field of view of Auger Detector.}
\end{figure*}

 The field of view of the Auger Observatory for photons ranges from declinations of -90$^{\circ}$ to 24.8$^{\circ}$. To increase our sensitivity to transients in those directions, we use a multi-flavor neutrino sample. The sample consists of two IceCube event selections. The first is an all-sky throughgoing tracks selection~\cite{pstracks}, which is predominately composed of muon neutrinos with well-reconstructed directions. These events are characterized by long "tracks" through the detector. The second selection is the cascades selection, which is primarily composed of neutral-current, $\nu_{e}^{CC}$ (charged-current), and $\nu_{\tau}^{CC}$ neutrino interactions that leave compact, spherical energy deposits in the detector. These events have less precise angular reconstruction. The cascades selection was previously utilized to detect the emission of high-energy neutrinos from the galactic plane at $4.5\sigma$~\cite{dnn}.  We remove overlapping events between these selections in data, choosing to leave the overlapping events in the cascade selection and removing them from the track selection (less than .1\% of tracks selection). \Cref{fig:effective area} shows the effective area of the tracks and cascade selections for neutrinos in the field of view of the Auger Detector. 
\subsection{Likelihood Construction}
\begin{figure*}[h!]
    \centering
    \begin{subfigure}[t]{0.59\textwidth}
        \centering
        \includegraphics[width=1\textwidth]{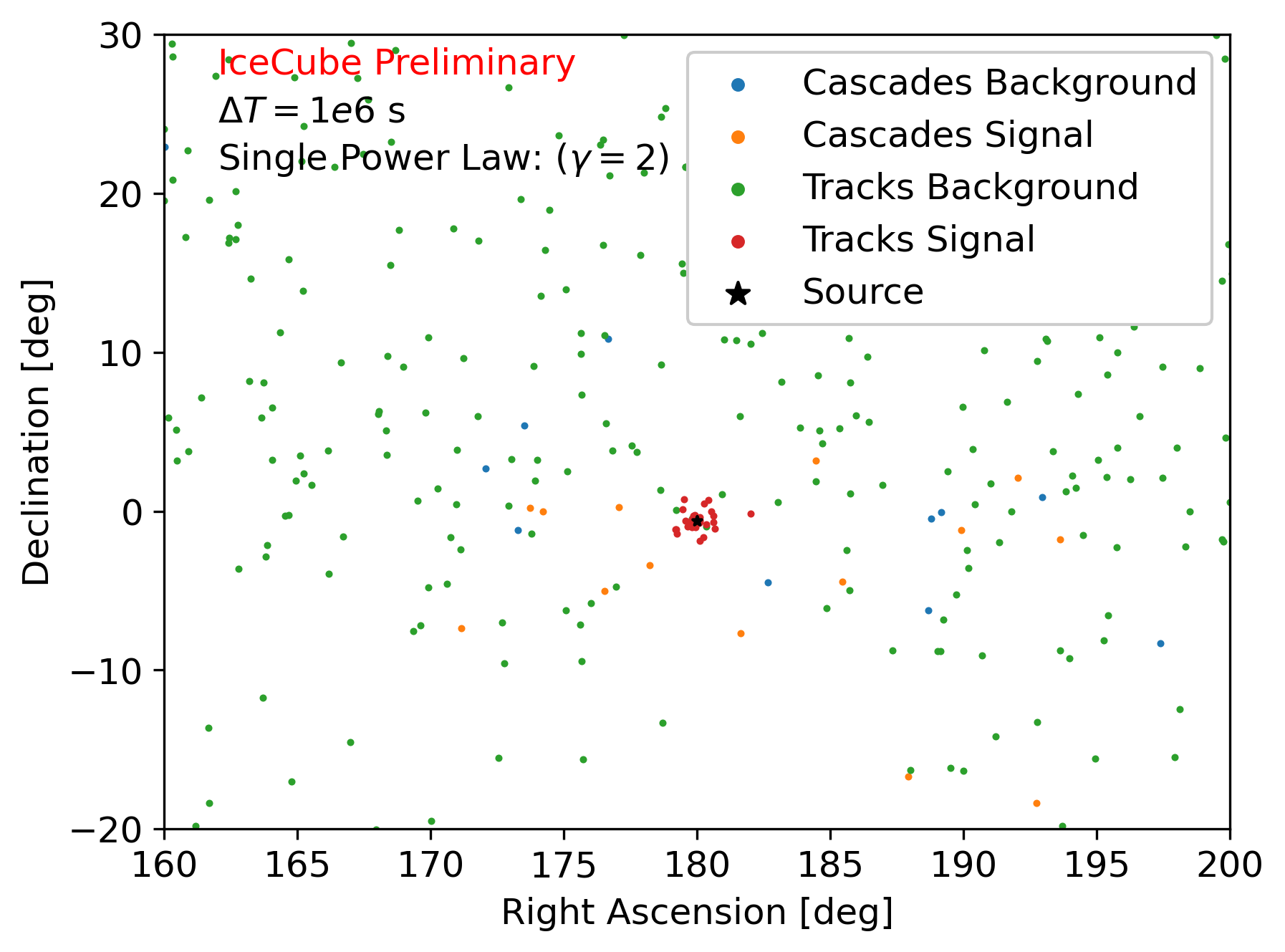}
        \captionsetup{skip=0pt}
        \caption{}
    \label{fig:spatialpdf}
    \end{subfigure}
    ~
    \begin{subfigure}[t]{0.39\textwidth}
        \centering
        \includegraphics[width=1\textwidth]{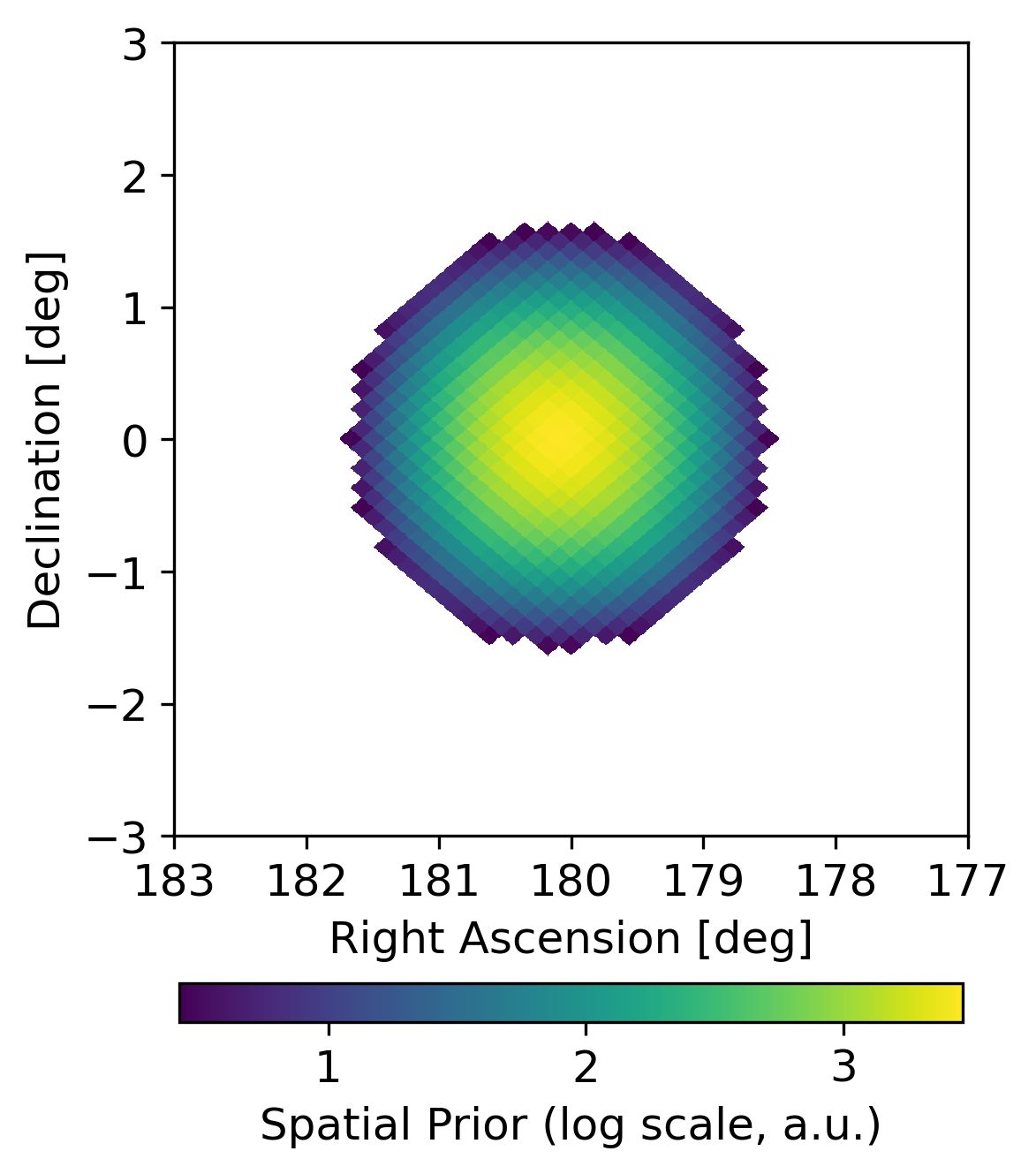}
        \captionsetup{skip=0pt}
        \caption{}
        \label{fig:spatial_prior}
    \end{subfigure}%
    \captionsetup{skip=0pt}
    \caption{\textbf{(a)} A single power law injection for a flare of a million seconds from an arbitrary direction in the Auger Detector's photon field of view. The longer flare is to increase the number of background events for visualization. Each point shows the reconstructed direction of a neutrino event. 50 total signal events where injected, and are split between tracks and cascades by the effective area of each selection. The spread in signal events comes from the precision with which events can be reconstructed. Cascades have a larger point spread function than tracks, so they have more spread. Background scrambled events from tracks and cascades are also shown, where the tracks selection has a higher event rate than cascades. \textbf{(b)} An example spatial prior map representing the uncertainty of a point source position with a gaussian $\sigma$ of 1 degree made by smoothing a HEALPix map using healpy \cite{healpix}. The spatial prior is masked for 99.9\% containment of the cumulative density function starting with the most probable position. The position on the sky is arbitrary. }
\end{figure*}

To perform this analysis, we construct an unbinned likelihood for a combined sample 
\begin{equation}
\mathcal{L}(\gamma , n_s) =  \prod_{j}^{M}  \frac{(n_s^j + n_b^j)^{N^j}e^{-(n_s^j+n_b^j)}}{N^j!}  \prod_{i \in j}^{N^j} \left[\frac{n_s^j}{n_s^j+n_b^j} \mathcal{S}_i^j(\delta_i^j, \sigma_i^j, \gamma = 2) + \frac{n_b^j}{n_s^j+n_b^j}\mathcal{B}_i^j(\delta_i^j) \right],
\end{equation}where $M$ is the number of samples and $j$ iterates over the samples, while $N^j$ is the total number of events in each sample occurring during the transient time window and $i$ iterates over those events. Additionally, $n_s$ is the fit number of signal events, and $n_b$ is the expected number of background events. $\mathcal{S}$ and $\mathcal{B}$ are the signal and background PDFs respectively, which are composed of energy, spatial, and time PDFs 
\begin{equation}
\mathcal{S}_i^j(\delta_i^j, \sigma_i^j, \gamma = 2) = \mathcal{S}_{\text{space}} \times \mathcal{S}_{\text{energy}} \times \mathcal{S}_{\text{time}}
\end{equation} 
as described in~\cite{method} and the background PDFs is constructed from
\begin{equation}
   \mathcal{B}_i^j(\delta_i^j) = \mathcal{B}_{\text{space}} \times \mathcal{B}_{\text{energy}} \times \mathcal{B}_{\text{time}}.
\end{equation}
where $\sigma_i$ is a per-event description of the point spread function (PSF) and $\delta_i$ is the declination. The signal energy PDF is generated assuming a $\gamma=2$ power law spectrum at the source and the background energy PDF is generated from data. The energy PDF describes the relative count of events at different energies for each declination. For example, a harder signal spectrum than background spectrum will result in a ratio above one at high energies when comparing the signal to background PDF. The signal spatial PDF is a circularized PSF described by a Kent distribution with a spread $\sigma_i$. When the spread is less than 7 degrees, the PSF is approximated by a Gaussian. The background spatial PDF is constructed using the data rate as a function of declination. Tracks have greater precision in reconstructed direction compared to cascades, so signal events will be clustered closer to the source, as depicted in \Cref{fig:spatialpdf}. 
The background spatial distribution is also depicted. The timing PDF is a boxed timing window centered on the photon candidate time with a total width of 20,000 seconds. Longer-time window estimates for LLGRBs and Long-GRBs have durations of up to 10,000 seconds~\cite{timingwindow}. The timing window is then extended to 10,000 seconds before the photon arrival time to allow for the case that the photon is emitted late in the flare. We construct a test statistic (TS) by maximizing the log-likelihood ratio 
\begin{equation}
\text{TS} = \max \left( -2 \ln \left[ \frac{\mathcal{L}(n_s = 0)}{\mathcal{L}(\hat{n_s})} \right] - 2 \ln \left[ \frac{1}{w(\Omega)} \right] \right),
\end{equation}
where a penalty is applied to the TS based on the location in the spatial prior map. The penalty term makes up the second component of the equation, where $w$ is probability that the source is at the location $\Omega$ on the sky over the best fit location of the photon candidate. We normalize the penalty factor $w$ so that it equals 1 at the most probable location. We compute the TS at each pixel in the spatial prior and pick the highest resulting TS. \Cref{fig:spatial_prior} depicts a spatial PDF that is a Gaussian with a spread ($\sigma$) of 1 degree. The location of the spatial PDF is arbitrarily chosen within the field of view of Auger Observatory. The size of the spatial prior was chosen to be similar to the size of the angular resolution of the hybrid photon candidates. The spatial prior is masked for 99.9\% containment of the cumulative PDF starting from the most significant pixel to the least. The significance of our results are computed from the TS.

\section{Sensitivities}
\subsection{Defining Sensitvities}
\begin{figure*}[h!]
    \centering
    \begin{subfigure}[t]{0.5\textwidth}
        \centering
        \includegraphics[width=1\textwidth]{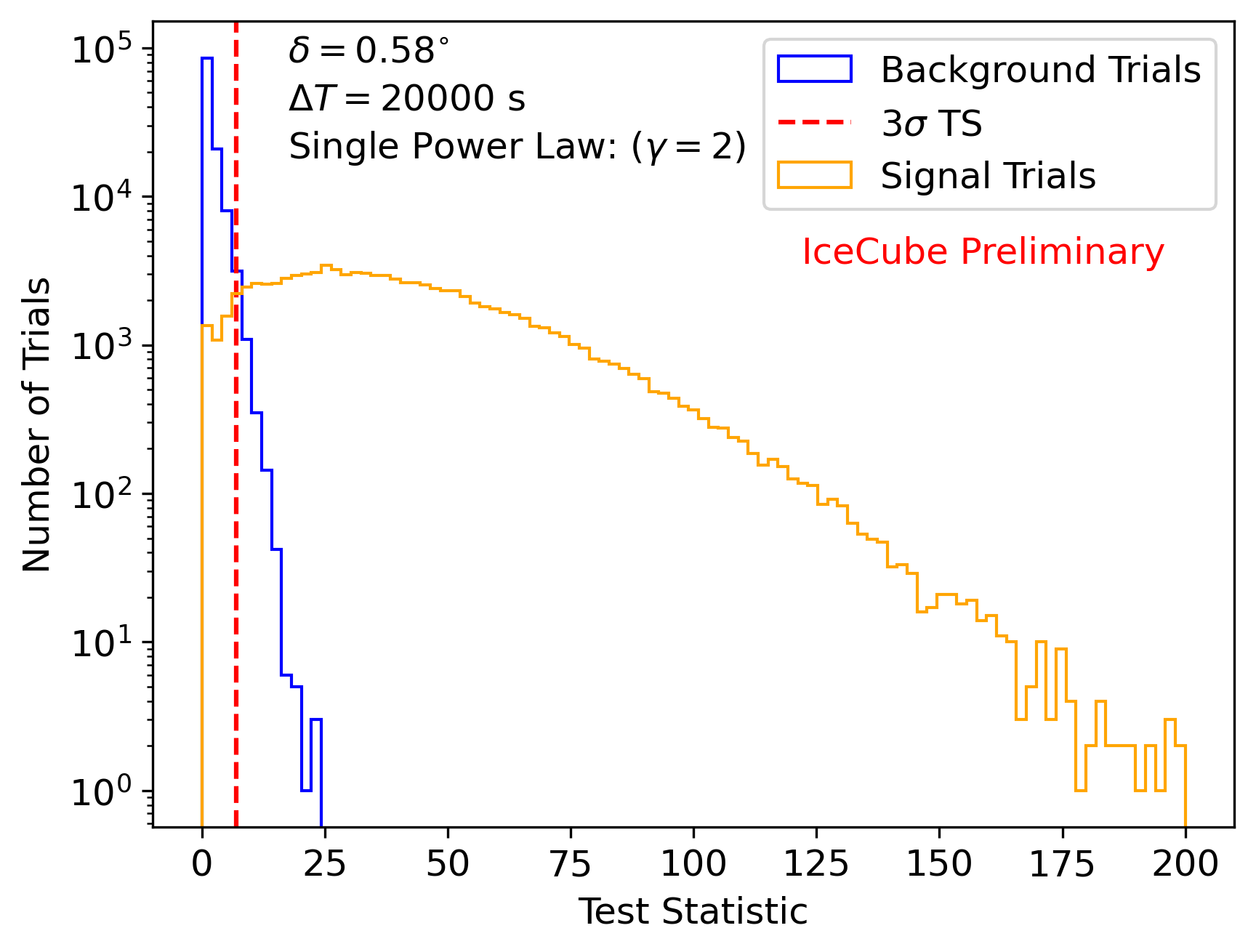}
        \captionsetup{skip=0pt}
        \caption{}
        \label{fig:bg_sig}
    \end{subfigure}%
    ~ 
    \begin{subfigure}[t]{0.5\textwidth}
        \centering
        \includegraphics[width=1\textwidth]{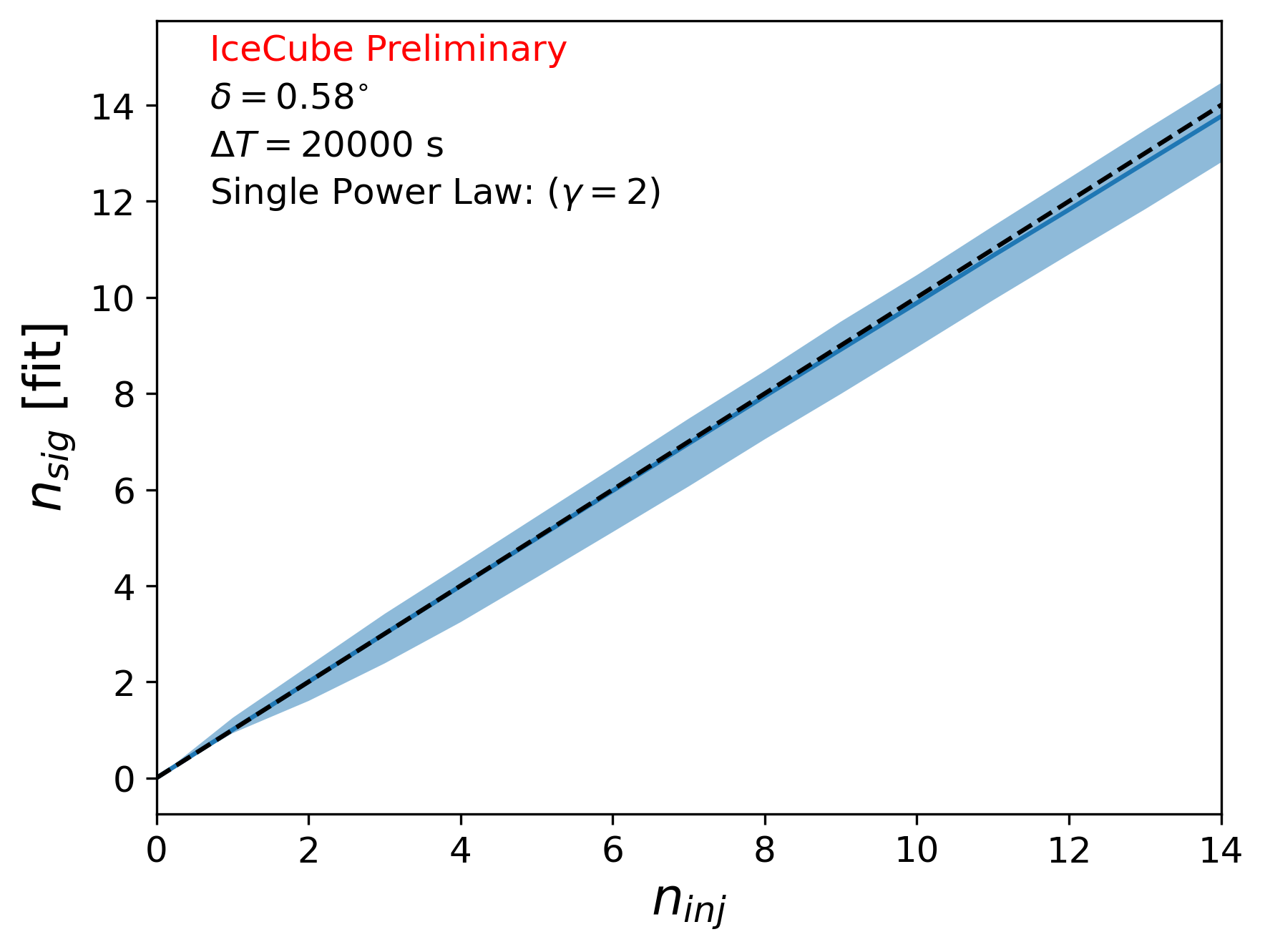}
        \captionsetup{skip=0pt}
        \caption{}
        \label{fig:bias}
    \end{subfigure}
    \label{fig:pointsource}
    \captionsetup{skip=0pt}
    \caption{\textbf{(a)} The background test statistic distribution is depicted in blue and the TS value corresponding to the $3\sigma$ sensitivity in red for a declination of $\delta = 0.58$. The orange line shows the test statistic distribution when injecting a mean of 3.42 signal events, which corresponds to the $3\sigma$ sensitivity. The fit is fixed at a spectral index of 2 and uses a boxed timing window. \textbf{(b)} An injection and recovery test at a declination of $\delta = 0.58$. In blue, the median and 1 sigma error bands for the fit $n_{sig}$ are shown for different $n_{inj}$. The dashed black line is the diagonal showing a ratio of 1 for fit $n_{sig}$ to $n_{inj}$. The fit is fixed at a spectral index of 2 and uses a boxed timing window.}
\end{figure*}

To construct our sensitivities, we perform background trials by randomly scrambling the neutrino data. For the scrambling procedure, we inject data events with Poisson fluctuations at the background rate from outside of the active flare time window. The arrival time of neutrino data is randomized within the flare timing window. We then evaluate the TS of each background trial using the unbinned maximum likelihood method. Over many trials, we produce the background TS distribution represented by the blue line in \Cref{fig:bg_sig}, which is an example picked at an arbitrary declination.  To find our sensitivity, we perform injection trials with Poisson fluctuations around a mean $n_s$ corresponding to a flux from a given source model increasing the strength until 90\% of the TS values are greater than the 3$\sigma$ TS value for the background TS distribution. The lowest fluence that meets the 90\% criteria is the 3$\sigma$ sensitivity. \Cref{fig:bg_sig} shows in orange the TS distribution for the number of signal events injected at the $3\sigma$ sensitivity. In addition, we demonstrate that the fit recovers the number of injected neutrinos over a wide range of injection strengths. The result of this process for a single point source is shown in \Cref{fig:bias}.

\subsection{Transient Sensitivities}

\begin{figure}[h]
\center
\includegraphics[width=.5\linewidth]{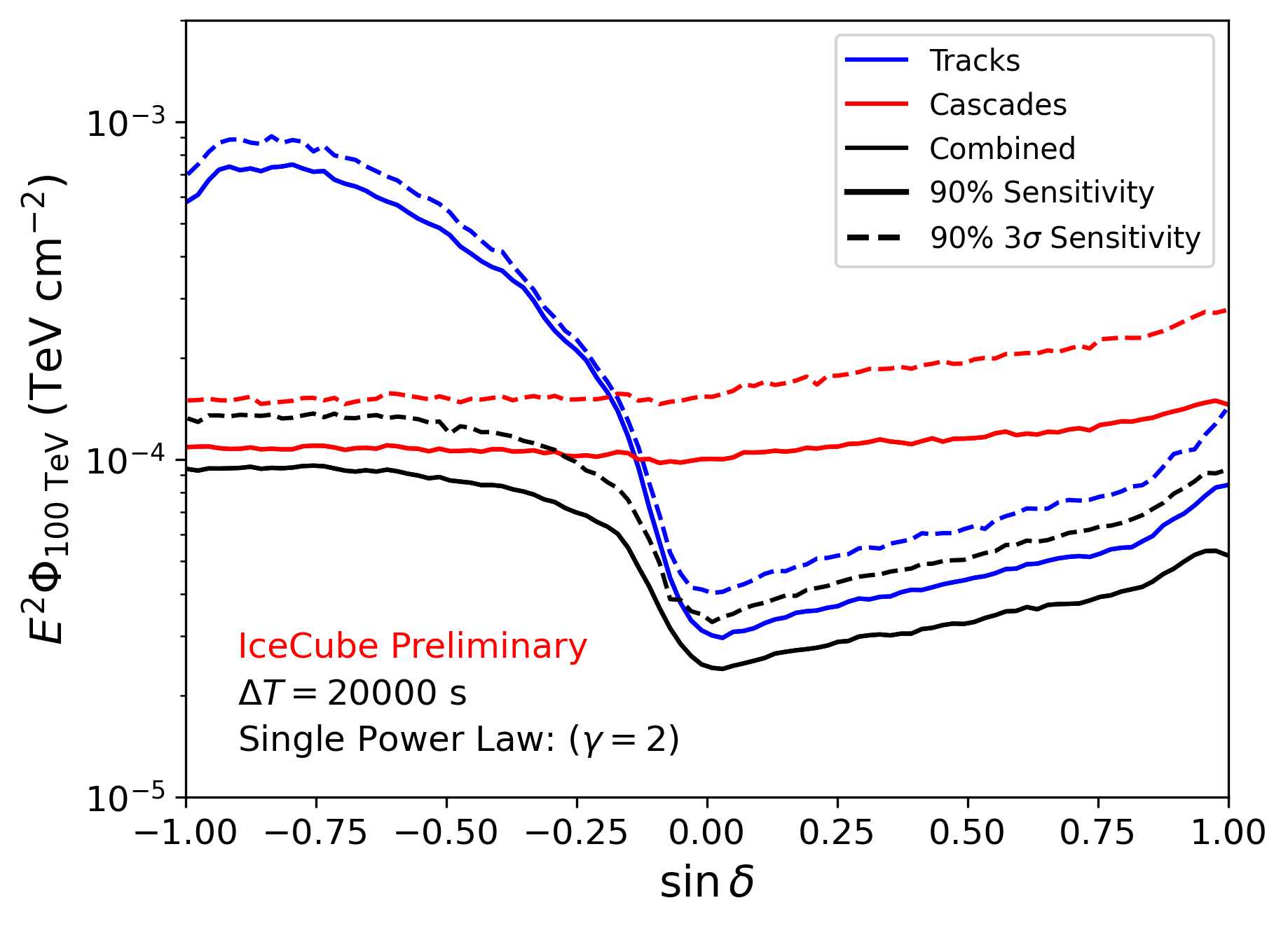}
\captionsetup{skip=0pt}
\caption{The transient sensitivities of the IceCube tracks and cascades sample are calculated using a point source likelihood for a source model with a single power law energy spectrum with a spectral index of 2, a flare duration of 20,000 seconds. The timing PDF uses a boxed timing window. In the northern sky and at the horizon, the sensitivity of the combined sample comes from tracks, whereas in the southern sky, where an extensive atmospheric muon background is present, the sensitivity of the combined sample is dominated by the cascades selection. }
\label{fig:dec}
\end{figure}

\begin{figure}[h]
\center
\includegraphics[width=.5\linewidth]{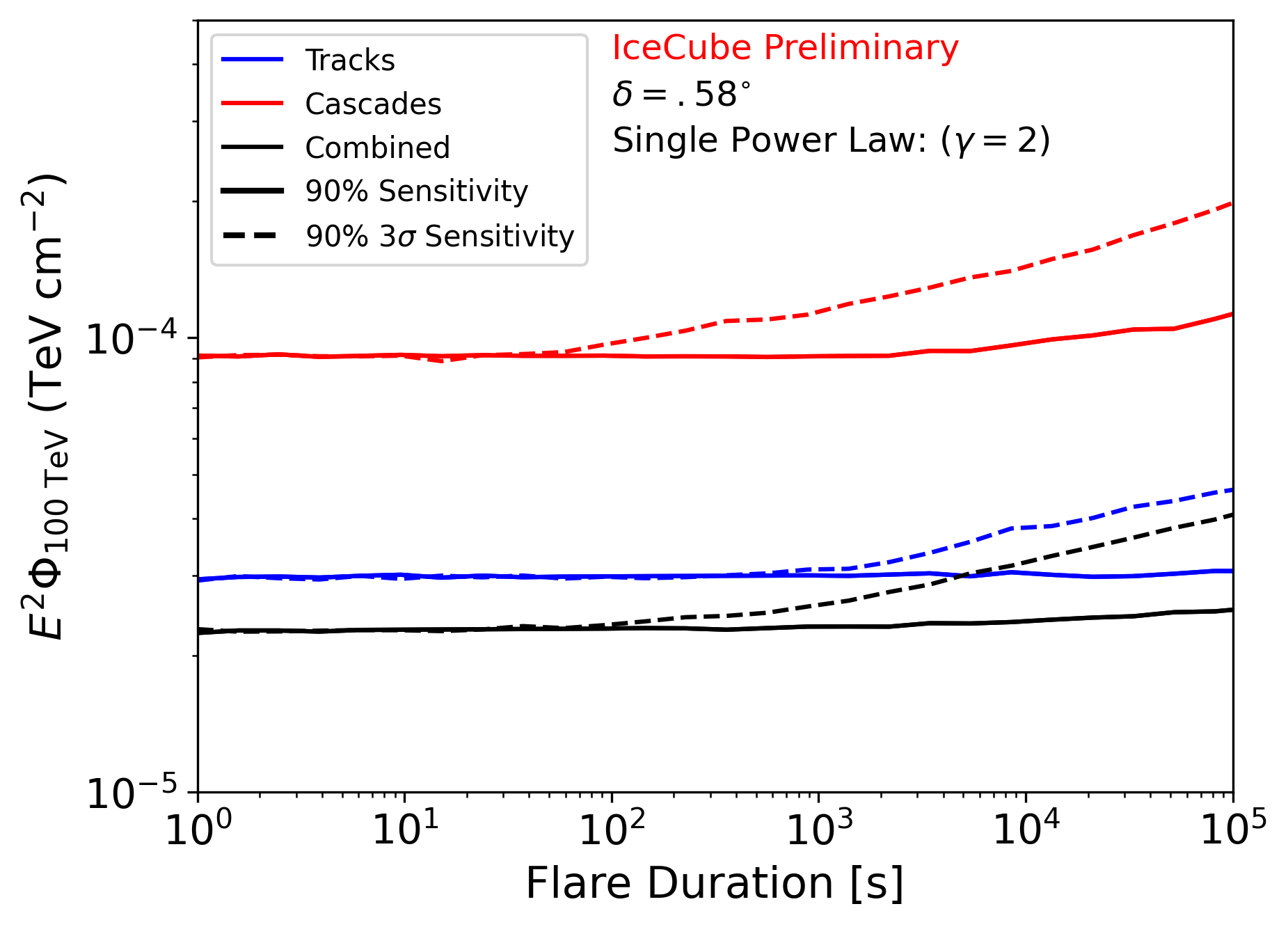}
\captionsetup{skip=0pt}
\caption{The sensitivities at different flare durations range from 1 to 100,000 seconds at a declination of $\delta = 0.58^{\circ}$. The injected source spectrum is a single power law with a spectral index of 2. The fit is fixed at a spectral index of 2 and uses a boxed timing window. The timing PDF uses a boxed timing window.}
\label{fig:timedep}
\end{figure}

With the multi-flavor neutrino sample and the EeV photon candidates, we will perform a search for UHE transient sources by performing the coincidence search described above. To evaluate the sensitivity of IceCube to transient neutrino sources, we evaluate tracks and cascades: both combined and individually, for a flare duration of 20,000 seconds centered on the photon candidate time at all declinations, as depicted in \Cref{fig:dec}. The effect of the choice of timing window for the sensitivities are depicted in \Cref{fig:timedep}. The 90\% sensitivity does not shift much with a larger timing window, but the 3$\sigma$ sensitivity worsens for flare durations larger than 100 seconds.  
\section{Conclusion}
We will conduct the first search for multi-messenger coincidences between EeV photon candidates and neutrinos to identify ultra-high-energy transient sources. We describe the unbinned likelihood maximization methodology used in the analysis, which involves searching for a significant detection of neutrinos in spatial, energy, and temporal coincidence with UHE photon candidates. A significant detection would provide new information, including the identification of photon candidates as true EeV photon showers rather than cosmic ray background, and also serve to identify astrophysical sources of ultra-high energy particles. 

\bibliographystyle{ICRC}
\setlength{\bibsep}{1pt}
\small{\bibliography{references}}

\clearpage

\section*{Full Author List: IceCube Collaboration}

\scriptsize
\noindent
R. Abbasi$^{16}$,
M. Ackermann$^{63}$,
J. Adams$^{17}$,
S. K. Agarwalla$^{39,\: {\rm a}}$,
J. A. Aguilar$^{10}$,
M. Ahlers$^{21}$,
J.M. Alameddine$^{22}$,
S. Ali$^{35}$,
N. M. Amin$^{43}$,
K. Andeen$^{41}$
% ...
\\
\\
% ...
$^{10}$ Universit{\'e} Libre de Bruxelles, Science Faculty CP230, B-1050 Brussels, Belgium \\
% ...
$^{16}$ Department of Physics, Loyola University Chicago, Chicago, IL 60660, USA \\
$^{17}$ Dept. of Physics and Astronomy, University of Canterbury, Private Bag 4800, Christchurch, New Zealand \\
% ...
$^{21}$ Niels Bohr Institute, University of Copenhagen, DK-2100 Copenhagen, Denmark \\
$^{22}$ Dept. of Physics, TU Dortmund University, D-44221 Dortmund, Germany \\
% ...
$^{35}$ Dept. of Physics and Astronomy, University of Kansas, Lawrence, KS 66045, USA \\
% ...
$^{39}$ Dept. of Physics and Wisconsin IceCube Particle Astrophysics Center, University of Wisconsin{\textemdash}Madison, Madison, WI 53706, USA \\
% ...
$^{41}$ Department of Physics, Marquette University, Milwaukee, WI 53201, USA \\
% ...
$^{43}$ Bartol Research Institute and Dept. of Physics and Astronomy, University of Delaware, Newark, DE 19716, USA \\
% ...
$^{63}$ Deutsches Elektronen-Synchrotron DESY, Platanenallee 6, D-15738 Zeuthen, Germany \\
$^{\rm a}$ also at Institute of Physics, Sachivalaya Marg, Sainik School Post, Bhubaneswar 751005, India \\
% ...

\subsection*{Acknowledgments}

\noindent
The authors gratefully acknowledge the support from the following agencies and institutions:
USA {\textendash} U.S. National Science Foundation-Office of Polar Programs,
U.S. National Science Foundation-Physics Division,
U.S. National Science Foundation-EPSCoR,
U.S. National Science Foundation-Office of Advanced Cyberinfrastructure,
Wisconsin Alumni Research Foundation,
Center for High Throughput Computing (CHTC) at the University of Wisconsin{\textendash}Madison,
Open Science Grid (OSG),
Partnership to Advance Throughput Computing (PATh),
Advanced Cyberinfrastructure Coordination Ecosystem: Services {\&} Support (ACCESS),
Frontera and Ranch computing project at the Texas Advanced Computing Center,
U.S. Department of Energy-National Energy Research Scientific Computing Center,
Particle astrophysics research computing center at the University of Maryland,
Institute for Cyber-Enabled Research at Michigan State University,
Astroparticle physics computational facility at Marquette University,
NVIDIA Corporation,
and Google Cloud Platform;
Belgium {\textendash} Funds for Scientific Research (FRS-FNRS and FWO),
FWO Odysseus and Big Science programmes,
and Belgian Federal Science Policy Office (Belspo);
Germany {\textendash} Bundesministerium f{\"u}r Forschung, Technologie und Raumfahrt (BMFTR),
Deutsche Forschungsgemeinschaft (DFG),
Helmholtz Alliance for Astroparticle Physics (HAP),
Initiative and Networking Fund of the Helmholtz Association,
Deutsches Elektronen Synchrotron (DESY),
and High Performance Computing cluster of the RWTH Aachen;
Sweden {\textendash} Swedish Research Council,
Swedish Polar Research Secretariat,
Swedish National Infrastructure for Computing (SNIC),
and Knut and Alice Wallenberg Foundation;
European Union {\textendash} EGI Advanced Computing for research;
Australia {\textendash} Australian Research Council;
Canada {\textendash} Natural Sciences and Engineering Research Council of Canada,
Calcul Qu{\'e}bec, Compute Ontario, Canada Foundation for Innovation, WestGrid, and Digital Research Alliance of Canada;
Denmark {\textendash} Villum Fonden, Carlsberg Foundation, and European Commission;
New Zealand {\textendash} Marsden Fund;
Japan {\textendash} Japan Society for Promotion of Science (JSPS)
and Institute for Global Prominent Research (IGPR) of Chiba University;
Korea {\textendash} National Research Foundation of Korea (NRF);
Switzerland {\textendash} Swiss National Science Foundation (SNSF).
\newpage
\section*{The Pierre Auger Collaboration}

{\footnotesize\setlength{\baselineskip}{10pt}
\noindent
\begin{wrapfigure}[11]{l}{0.12\linewidth}
\vspace{-4pt}
\includegraphics[width=0.98\linewidth]{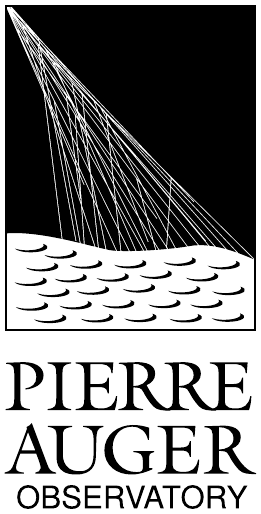}
\end{wrapfigure}
\begin{sloppypar}\noindent
A.~Abdul Halim$^{13}$,
P.~Abreu$^{70}$,
M.~Aglietta$^{53,51}$,
I.~Allekotte$^{1}$,
K.~Almeida Cheminant$^{78,77}$,
A.~Almela$^{7,12}$,
R.~Aloisio$^{44,45}$,
J.~Alvarez-Mu\~niz$^{76}$,
A.~Ambrosone$^{44}$,
J.~Ammerman Yebra$^{76}$
% ...

\end{sloppypar}
\begin{center}
\end{center}

\vspace{1ex}
\begin{description}[labelsep=0.2em,align=right,labelwidth=0.7em,labelindent=0em,leftmargin=2em,noitemsep,before={\renewcommand\makelabel[1]{##1 }}]
\item[$^{1}$] Centro At\'omico Bariloche and Instituto Balseiro (CNEA-UNCuyo-CONICET), San Carlos de Bariloche, Argentina
\item[$^{2}$] Departamento de F\'\i{}sica and Departamento de Ciencias de la Atm\'osfera y los Oc\'eanos, FCEyN, Universidad de Buenos Aires and CONICET, Buenos Aires, Argentina
\item[$^{3}$] IFLP, Universidad Nacional de La Plata and CONICET, La Plata, Argentina
\item[$^{4}$] Instituto de Astronom\'\i{}a y F\'\i{}sica del Espacio (IAFE, CONICET-UBA), Buenos Aires, Argentina
\item[$^{5}$] Instituto de F\'\i{}sica de Rosario (IFIR) -- CONICET/U.N.R.\ and Facultad de Ciencias Bioqu\'\i{}micas y Farmac\'euticas U.N.R., Rosario, Argentina
\item[$^{6}$] Instituto de Tecnolog\'\i{}as en Detecci\'on y Astropart\'\i{}culas (CNEA, CONICET, UNSAM), and Universidad Tecnol\'ogica Nacional -- Facultad Regional Mendoza (CONICET/CNEA), Mendoza, Argentina
\item[$^{7}$] Instituto de Tecnolog\'\i{}as en Detecci\'on y Astropart\'\i{}culas (CNEA, CONICET, UNSAM), Buenos Aires, Argentina
\item[$^{8}$] International Center of Advanced Studies and Instituto de Ciencias F\'\i{}sicas, ECyT-UNSAM and CONICET, Campus Miguelete -- San Mart\'\i{}n, Buenos Aires, Argentina
\item[$^{9}$] Laboratorio Atm\'osfera -- Departamento de Investigaciones en L\'aseres y sus Aplicaciones -- UNIDEF (CITEDEF-CONICET), Argentina
\item[$^{10}$] Observatorio Pierre Auger, Malarg\"ue, Argentina
\item[$^{11}$] Observatorio Pierre Auger and Comisi\'on Nacional de Energ\'\i{}a At\'omica, Malarg\"ue, Argentina
\item[$^{12}$] Universidad Tecnol\'ogica Nacional -- Facultad Regional Buenos Aires, Buenos Aires, Argentina
\item[$^{13}$] University of Adelaide, Adelaide, S.A., Australia
\item[$^{14}$] Universit\'e Libre de Bruxelles (ULB), Brussels, Belgium
\item[$^{15}$] Vrije Universiteit Brussels, Brussels, Belgium
\item[$^{16}$] Centro Brasileiro de Pesquisas Fisicas, Rio de Janeiro, RJ, Brazil
\item[$^{17}$] Centro Federal de Educa\c{c}\~ao Tecnol\'ogica Celso Suckow da Fonseca, Petropolis, Brazil
\item[$^{18}$] Instituto Federal de Educa\c{c}\~ao, Ci\^encia e Tecnologia do Rio de Janeiro (IFRJ), Brazil
\item[$^{19}$] Universidade de S\~ao Paulo, Escola de Engenharia de Lorena, Lorena, SP, Brazil
\item[$^{20}$] Universidade de S\~ao Paulo, Instituto de F\'\i{}sica de S\~ao Carlos, S\~ao Carlos, SP, Brazil
\item[$^{21}$] Universidade de S\~ao Paulo, Instituto de F\'\i{}sica, S\~ao Paulo, SP, Brazil
\item[$^{22}$] Universidade Estadual de Campinas (UNICAMP), IFGW, Campinas, SP, Brazil
\item[$^{23}$] Universidade Estadual de Feira de Santana, Feira de Santana, Brazil
\item[$^{24}$] Universidade Federal de Campina Grande, Centro de Ciencias e Tecnologia, Campina Grande, Brazil
\item[$^{25}$] Universidade Federal do ABC, Santo Andr\'e, SP, Brazil
\item[$^{26}$] Universidade Federal do Paran\'a, Setor Palotina, Palotina, Brazil
\item[$^{27}$] Universidade Federal do Rio de Janeiro, Instituto de F\'\i{}sica, Rio de Janeiro, RJ, Brazil
\item[$^{28}$] Universidad de Medell\'\i{}n, Medell\'\i{}n, Colombia
\item[$^{29}$] Universidad Industrial de Santander, Bucaramanga, Colombia
\item[$^{30}$] Charles University, Faculty of Mathematics and Physics, Institute of Particle and Nuclear Physics, Prague, Czech Republic
\item[$^{31}$] Institute of Physics of the Czech Academy of Sciences, Prague, Czech Republic
\item[$^{32}$] Palacky University, Olomouc, Czech Republic
\item[$^{33}$] CNRS/IN2P3, IJCLab, Universit\'e Paris-Saclay, Orsay, France
\item[$^{34}$] Laboratoire de Physique Nucl\'eaire et de Hautes Energies (LPNHE), Sorbonne Universit\'e, Universit\'e de Paris, CNRS-IN2P3, Paris, France
\item[$^{35}$] Univ.\ Grenoble Alpes, CNRS, Grenoble Institute of Engineering Univ.\ Grenoble Alpes, LPSC-IN2P3, 38000 Grenoble, France
\item[$^{36}$] Universit\'e Paris-Saclay, CNRS/IN2P3, IJCLab, Orsay, France
\item[$^{37}$] Bergische Universit\"at Wuppertal, Department of Physics, Wuppertal, Germany
\item[$^{38}$] Karlsruhe Institute of Technology (KIT), Institute for Experimental Particle Physics, Karlsruhe, Germany
\item[$^{39}$] Karlsruhe Institute of Technology (KIT), Institut f\"ur Prozessdatenverarbeitung und Elektronik, Karlsruhe, Germany
\item[$^{40}$] Karlsruhe Institute of Technology (KIT), Institute for Astroparticle Physics, Karlsruhe, Germany
\item[$^{41}$] RWTH Aachen University, III.\ Physikalisches Institut A, Aachen, Germany
\item[$^{42}$] Universit\"at Hamburg, II.\ Institut f\"ur Theoretische Physik, Hamburg, Germany
\item[$^{43}$] Universit\"at Siegen, Department Physik -- Experimentelle Teilchenphysik, Siegen, Germany
\item[$^{44}$] Gran Sasso Science Institute, L'Aquila, Italy
\item[$^{45}$] INFN Laboratori Nazionali del Gran Sasso, Assergi (L'Aquila), Italy
\item[$^{46}$] INFN, Sezione di Catania, Catania, Italy
\item[$^{47}$] INFN, Sezione di Lecce, Lecce, Italy
\item[$^{48}$] INFN, Sezione di Milano, Milano, Italy
\item[$^{49}$] INFN, Sezione di Napoli, Napoli, Italy
\item[$^{50}$] INFN, Sezione di Roma ``Tor Vergata'', Roma, Italy
\item[$^{51}$] INFN, Sezione di Torino, Torino, Italy
\item[$^{52}$] Istituto di Astrofisica Spaziale e Fisica Cosmica di Palermo (INAF), Palermo, Italy
\item[$^{53}$] Osservatorio Astrofisico di Torino (INAF), Torino, Italy
\item[$^{54}$] Politecnico di Milano, Dipartimento di Scienze e Tecnologie Aerospaziali , Milano, Italy
\item[$^{55}$] Universit\`a del Salento, Dipartimento di Matematica e Fisica ``E.\ De Giorgi'', Lecce, Italy
\item[$^{56}$] Universit\`a dell'Aquila, Dipartimento di Scienze Fisiche e Chimiche, L'Aquila, Italy
\item[$^{57}$] Universit\`a di Catania, Dipartimento di Fisica e Astronomia ``Ettore Majorana``, Catania, Italy
\item[$^{58}$] Universit\`a di Milano, Dipartimento di Fisica, Milano, Italy
\item[$^{59}$] Universit\`a di Napoli ``Federico II'', Dipartimento di Fisica ``Ettore Pancini'', Napoli, Italy
\item[$^{60}$] Universit\`a di Palermo, Dipartimento di Fisica e Chimica ''E.\ Segr\`e'', Palermo, Italy
\item[$^{61}$] Universit\`a di Roma ``Tor Vergata'', Dipartimento di Fisica, Roma, Italy
\item[$^{62}$] Universit\`a Torino, Dipartimento di Fisica, Torino, Italy
\item[$^{63}$] Benem\'erita Universidad Aut\'onoma de Puebla, Puebla, M\'exico
\item[$^{64}$] Unidad Profesional Interdisciplinaria en Ingenier\'\i{}a y Tecnolog\'\i{}as Avanzadas del Instituto Polit\'ecnico Nacional (UPIITA-IPN), M\'exico, D.F., M\'exico
\item[$^{65}$] Universidad Aut\'onoma de Chiapas, Tuxtla Guti\'errez, Chiapas, M\'exico
\item[$^{66}$] Universidad Michoacana de San Nicol\'as de Hidalgo, Morelia, Michoac\'an, M\'exico
\item[$^{67}$] Universidad Nacional Aut\'onoma de M\'exico, M\'exico, D.F., M\'exico
\item[$^{68}$] Institute of Nuclear Physics PAN, Krakow, Poland
\item[$^{69}$] University of \L{}\'od\'z, Faculty of High-Energy Astrophysics,\L{}\'od\'z, Poland
\item[$^{70}$] Laborat\'orio de Instrumenta\c{c}\~ao e F\'\i{}sica Experimental de Part\'\i{}culas -- LIP and Instituto Superior T\'ecnico -- IST, Universidade de Lisboa -- UL, Lisboa, Portugal
\item[$^{71}$] ``Horia Hulubei'' National Institute for Physics and Nuclear Engineering, Bucharest-Magurele, Romania
\item[$^{72}$] Institute of Space Science, Bucharest-Magurele, Romania
\item[$^{73}$] Center for Astrophysics and Cosmology (CAC), University of Nova Gorica, Nova Gorica, Slovenia
\item[$^{74}$] Experimental Particle Physics Department, J.\ Stefan Institute, Ljubljana, Slovenia
\item[$^{75}$] Universidad de Granada and C.A.F.P.E., Granada, Spain
\item[$^{76}$] Instituto Galego de F\'\i{}sica de Altas Enerx\'\i{}as (IGFAE), Universidade de Santiago de Compostela, Santiago de Compostela, Spain
\item[$^{77}$] IMAPP, Radboud University Nijmegen, Nijmegen, The Netherlands
\item[$^{78}$] Nationaal Instituut voor Kernfysica en Hoge Energie Fysica (NIKHEF), Science Park, Amsterdam, The Netherlands
\item[$^{79}$] Stichting Astronomisch Onderzoek in Nederland (ASTRON), Dwingeloo, The Netherlands
\item[$^{80}$] Universiteit van Amsterdam, Faculty of Science, Amsterdam, The Netherlands
\item[$^{81}$] Case Western Reserve University, Cleveland, OH, USA
\item[$^{82}$] Colorado School of Mines, Golden, CO, USA
\item[$^{83}$] Department of Physics and Astronomy, Lehman College, City University of New York, Bronx, NY, USA
\item[$^{84}$] Michigan Technological University, Houghton, MI, USA
\item[$^{85}$] New York University, New York, NY, USA
\item[$^{86}$] University of Chicago, Enrico Fermi Institute, Chicago, IL, USA
\item[$^{87}$] University of Delaware, Department of Physics and Astronomy, Bartol Research Institute, Newark, DE, USA
\item[] -----
\item[$^{a}$] Max-Planck-Institut f\"ur Radioastronomie, Bonn, Germany
\item[$^{b}$] also at Kapteyn Institute, University of Groningen, Groningen, The Netherlands
\item[$^{c}$] School of Physics and Astronomy, University of Leeds, Leeds, United Kingdom
\item[$^{d}$] Fermi National Accelerator Laboratory, Fermilab, Batavia, IL, USA
\item[$^{e}$] Pennsylvania State University, University Park, PA, USA
\item[$^{f}$] Colorado State University, Fort Collins, CO, USA
\item[$^{g}$] Louisiana State University, Baton Rouge, LA, USA
\item[$^{h}$] now at Graduate School of Science, Osaka Metropolitan University, Osaka, Japan
\item[$^{i}$] Institut universitaire de France (IUF), France
\item[$^{j}$] now at Technische Universit\"at Dortmund and Ruhr-Universit\"at Bochum, Dortmund and Bochum, Germany
\end{description}

\section*{Acknowledgments}

\begin{sloppypar}
The successful installation, commissioning, and operation of the Pierre
Auger Observatory would not have been possible without the strong
commitment and effort from the technical and administrative staff in
Malarg\"ue. We are very grateful to the following agencies and
organizations for financial support:
\end{sloppypar}

\begin{sloppypar}
Argentina -- Comisi\'on Nacional de Energ\'\i{}a At\'omica; Agencia Nacional de
Promoci\'on Cient\'\i{}fica y Tecnol\'ogica (ANPCyT); Consejo Nacional de
Investigaciones Cient\'\i{}ficas y T\'ecnicas (CONICET); Gobierno de la
Provincia de Mendoza; Municipalidad de Malarg\"ue; NDM Holdings and Valle
Las Le\~nas; in gratitude for their continuing cooperation over land
access; Australia -- the Australian Research Council; Belgium -- Fonds
de la Recherche Scientifique (FNRS); Research Foundation Flanders (FWO),
Marie Curie Action of the European Union Grant No.~101107047; Brazil --
Conselho Nacional de Desenvolvimento Cient\'\i{}fico e Tecnol\'ogico (CNPq);
Financiadora de Estudos e Projetos (FINEP); Funda\c{c}\~ao de Amparo \`a
Pesquisa do Estado de Rio de Janeiro (FAPERJ); S\~ao Paulo Research
Foundation (FAPESP) Grants No.~2019/10151-2, No.~2010/07359-6 and
No.~1999/05404-3; Minist\'erio da Ci\^encia, Tecnologia, Inova\c{c}\~oes e
Comunica\c{c}\~oes (MCTIC); Czech Republic -- GACR 24-13049S, CAS LQ100102401,
MEYS LM2023032, CZ.02.1.01/0.0/0.0/16{\textunderscore}013/0001402,
CZ.02.1.01/0.0/0.0/18{\textunderscore}046/0016010 and
CZ.02.1.01/0.0/0.0/17{\textunderscore}049/0008422 and CZ.02.01.01/00/22{\textunderscore}008/0004632;
France -- Centre de Calcul IN2P3/CNRS; Centre National de la Recherche
Scientifique (CNRS); Conseil R\'egional Ile-de-France; D\'epartement
Physique Nucl\'eaire et Corpusculaire (PNC-IN2P3/CNRS); D\'epartement
Sciences de l'Univers (SDU-INSU/CNRS); Institut Lagrange de Paris (ILP)
Grant No.~LABEX ANR-10-LABX-63 within the Investissements d'Avenir
Programme Grant No.~ANR-11-IDEX-0004-02; Germany -- Bundesministerium
f\"ur Bildung und Forschung (BMBF); Deutsche Forschungsgemeinschaft (DFG);
Finanzministerium Baden-W\"urttemberg; Helmholtz Alliance for
Astroparticle Physics (HAP); Helmholtz-Gemeinschaft Deutscher
Forschungszentren (HGF); Ministerium f\"ur Kultur und Wissenschaft des
Landes Nordrhein-Westfalen; Ministerium f\"ur Wissenschaft, Forschung und
Kunst des Landes Baden-W\"urttemberg; Italy -- Istituto Nazionale di
Fisica Nucleare (INFN); Istituto Nazionale di Astrofisica (INAF);
Ministero dell'Universit\`a e della Ricerca (MUR); CETEMPS Center of
Excellence; Ministero degli Affari Esteri (MAE), ICSC Centro Nazionale
di Ricerca in High Performance Computing, Big Data and Quantum
Computing, funded by European Union NextGenerationEU, reference code
CN{\textunderscore}00000013; M\'exico -- Consejo Nacional de Ciencia y Tecnolog\'\i{}a
(CONACYT) No.~167733; Universidad Nacional Aut\'onoma de M\'exico (UNAM);
PAPIIT DGAPA-UNAM; The Netherlands -- Ministry of Education, Culture and
Science; Netherlands Organisation for Scientific Research (NWO); Dutch
national e-infrastructure with the support of SURF Cooperative; Poland
-- Ministry of Education and Science, grants No.~DIR/WK/2018/11 and
2022/WK/12; National Science Centre, grants No.~2016/22/M/ST9/00198,
2016/23/B/ST9/01635, 2020/39/B/ST9/01398, and 2022/45/B/ST9/02163;
Portugal -- Portuguese national funds and FEDER funds within Programa
Operacional Factores de Competitividade through Funda\c{c}\~ao para a Ci\^encia
e a Tecnologia (COMPETE); Romania -- Ministry of Research, Innovation
and Digitization, CNCS-UEFISCDI, contract no.~30N/2023 under Romanian
National Core Program LAPLAS VII, grant no.~PN 23 21 01 02 and project
number PN-III-P1-1.1-TE-2021-0924/TE57/2022, within PNCDI III; Slovenia
-- Slovenian Research Agency, grants P1-0031, P1-0385, I0-0033, N1-0111;
Spain -- Ministerio de Ciencia e Innovaci\'on/Agencia Estatal de
Investigaci\'on (PID2019-105544GB-I00, PID2022-140510NB-I00 and
RYC2019-027017-I), Xunta de Galicia (CIGUS Network of Research Centers,
Consolidaci\'on 2021 GRC GI-2033, ED431C-2021/22 and ED431F-2022/15),
Junta de Andaluc\'\i{}a (SOMM17/6104/UGR and P18-FR-4314), and the European
Union (Marie Sklodowska-Curie 101065027 and ERDF); USA -- Department of
Energy, Contracts No.~DE-AC02-07CH11359, No.~DE-FR02-04ER41300,
No.~DE-FG02-99ER41107 and No.~DE-SC0011689; National Science Foundation,
Grant No.~0450696, and NSF-2013199; The Grainger Foundation; Marie
Curie-IRSES/EPLANET; European Particle Physics Latin American Network;
and UNESCO.
\end{sloppypar}

}
\end{document}